\documentclass[submitting,groupedaddress]{nst}

\usepackage{subfigure,dcolumn}
\usepackage[T2A,T1]{fontenc}
\usepackage[russian,english]{babel}

\usepackage{listings}
\usepackage[colorlinks=true,urlcolor=blue]{hyperref}
            
\usepackage{lineno}
\renewcommand{\linenumbers}{}
\nolinenumbers
\begin{document}
%%%%%%%%%%%%%%%%%%%%%%%%%%%%%%%%%%%%%%%%%%%%%%%%%%%%%%%%%%%%%%%%%%

\title{A Large-Volume GAGG(Ce) Detector Module for In-Beam $\gamma$-Ray Spectroscopy}

\author{Jun-Qi~Li}
\author{Ye-Lei~Sun}
\email{sunyelei@buaa.edu.cn}
\author{Ming-Yuan~Sun}
\author{Dan-Yang~Li}
\author{Hao~Wu}
\author{Bao-Hua~Sun}
\email{bhsun@buaa.edu.cn}

\affiliation{
School of Physics, Beihang University,
Beijing 100191, China
}

\author{Hong-Na~Liu}
\email{hongna.liu@bnu.edu.cn}
\author{Jun~Su}
\author{Xin~Chen}
\author{Xin-Zhi~Jiang}
\author{Yi-Tong~Huang}
\author{Yong-Liang~Jiang}
\author{Tuo-Qi~Liu}
\author{Yun-Long~Lu}
\author{Lin~Wang}
\author{Yu-Yang~Yu}

\affiliation{
School of Physics and Astronomy, Beijing Normal University,
Beijing 100875, China
}

\begin{abstract}
We report on the development and characterization of large-volume GAGG(Ce) 
scintillation detector modules equipped with silicon photomultiplier (SiPM) readout.
Each module houses four large-volume GAGG(Ce) crystals within a compact thin carbon-fiber enclosure, minimizing the passive material budget and enabling the construction of high-efficiency, large-solid-angle detector arrays for in-beam $\gamma$-ray spectroscopy. 
The performance of 10 GAGG modules (40 detectors), including the energy linearity, energy and timing resolution, and intrinsic full-energy peak efficiency, was evaluated using standard $\gamma$-ray sources and $^{14}$N($p$, $\gamma$)$^{15}$O reactions, covering a gamma-ray energy range from about 0.3 MeV to 8 MeV. Non-linear energy response across the measured energy range was observed due to the non-linear response of the SiPM. The detector modules exhibit good energy resolution (FWHM below 5\% at 1.33 MeV), good timing resolution (FWHM $\sim$500 ps at $\sim$1 MeV), and intrinsic full-energy peak efficiencies in good agreement with Geant4 simulations, making the detector modules a promising solution for 
next-generation high efficiency in-beam $\gamma$-ray experiments at RI-beam facilities.
 
\end{abstract}


\maketitle

\section{Introduction}

In-beam $\gamma$-ray spectroscopy is a key experimental technique in nuclear physics. In this approach, heavy-ion beams are utilized to bombard target nuclei, populating nuclear excited states. By measuring the energy, intensity, and timing information of the characteristic $\gamma$-rays emitted during the de-excitation process, detailed information on nuclear level schemes and transition properties can be systematically investigated. It is a key approach to study the nuclear deformation, shell evolution, and the nature of nuclear forces, particularly in exotic nuclei far from the stable line~\cite{Pieter2012, Otsuka2020, Ye2025}.

The core instrument for in-beam $\gamma$-ray spectroscopy is the $\gamma$-ray spectrometer, the development of which has followed two complementary directions. 
One direction aims to achieve the highest possible energy resolution using germanium detectors, particularly $\gamma$-ray tracking arrays~\cite{AGATA2012,GRETINA2013}, 
while the other focuses on developing high-efficiency and cost-effective scintillator-based detector arrays while maintaining acceptable energy resolution.
Due to the limited intensities of radioactive ion beams (RIBs), the development of $\gamma$-ray spectrometers with high detection efficiency is indeed essential.
Representative high-detection efficiency $\gamma$-ray spectrometers include DALI2 based on NaI(Tl) scintillators at RIKEN~\cite{Takeuchi2014}, CAESAR based on CsI(Na) scintillators at NSCL~\cite{CAESAR_GILLESPIE2026171217} and CALIFA based on CsI(Tl) scintillators at GSI~\cite{Cortina2014}, among others.

 \begin{figure*}[t!]
   \setlength{\abovecaptionskip}{0pt}
   \setlength{\belowcaptionskip}{0pt}
   \centering
   \includegraphics[width=\textwidth]{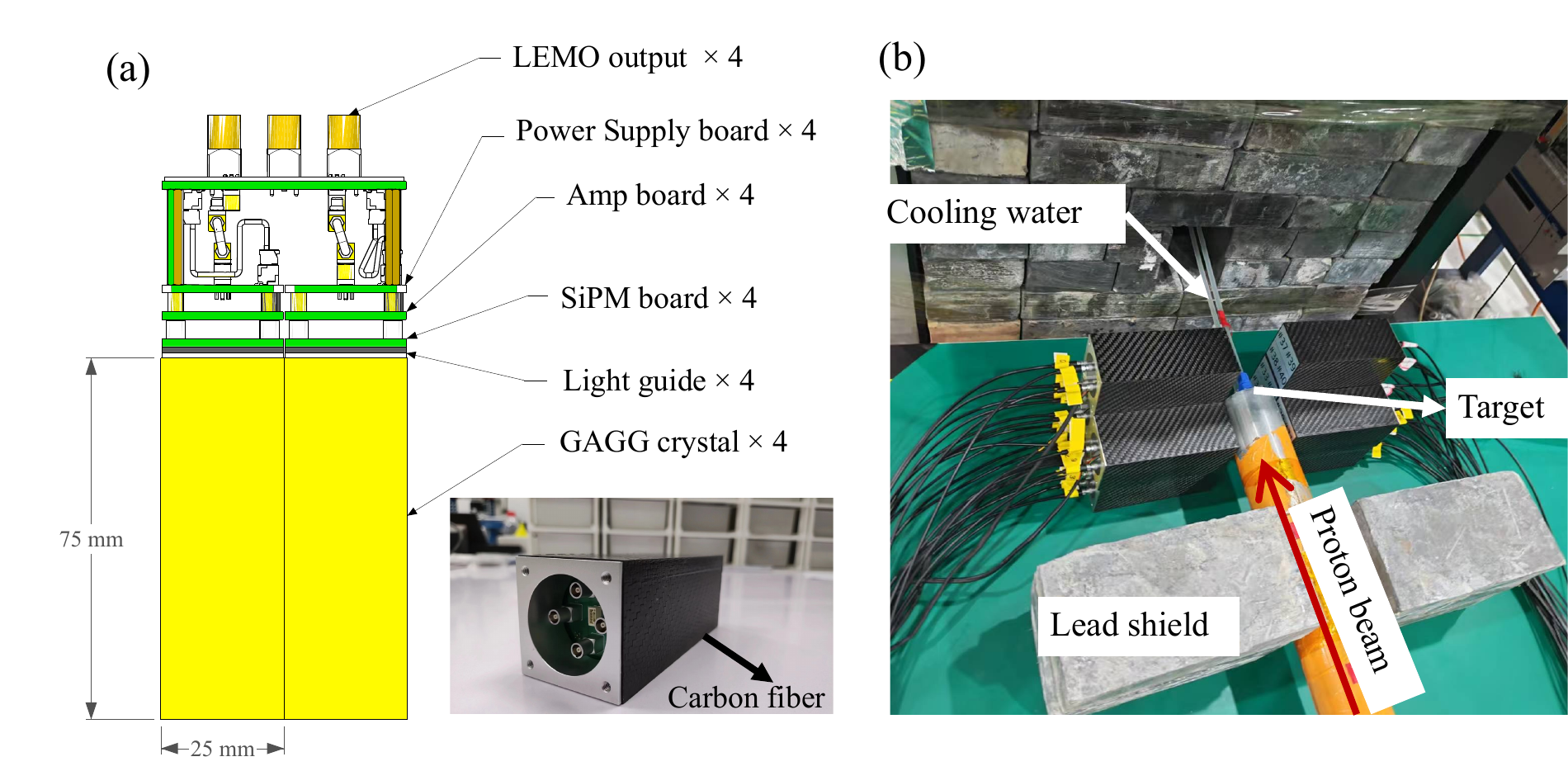}
    \caption{(a) Structural layout and front-end signal readout design of the GAGG detector module. (b) Experimental setup of the detector test using the $^{14}$N($p$, $\gamma$)$^{15}$O reaction at HINEG.}
    \label{fig:fig1}
\end{figure*}

While these scintillator-based spectrometers have 
enabled a broad range of studies with exotic beams, further improvements in energy resolution, timing performance, and compactness are 
essential for next-generation experiments to fully exploit the intense RIBs available at the RI beam facilities and to enable spectroscopic studies of the most exotic nuclei near the drip lines.
Cerium-doped Gadolinium Aluminum Gallium Garnet (GAGG(Ce)) is a modern inorganic scintillator developed in the early 2010s~\cite{Kamada2011,Kamada2012,Kamada2014,Kamada2016,Zhu2020} and has attracted significant attention in nuclear and particle physics for $\gamma$-ray, charged-particle, and neutron detection~\cite{Kobayashi2012,Iwanowska2013,Taggart2019,Furuno2021,Martinazzoli2021,Tanaka2023}, in medical physics applications~\cite{Schneider2015,Gerasymov2020}, as well as in space-based $\gamma$-ray and neutrino detection~\cite{Yoneyama2018,Wen2019,Solomey2023,Pallu2024,Doty2025}. 
It exhibits several advantageous properties compared with traditional scintillators 
such as NaI(Tl) and CsI(Tl), including high light yield, high density and effective atomic number, fast scintillation response, and non-hygroscopicity. Compared with 
rare-earth halide scintillators such as LaBr$_3$(Ce)~\cite{Loef2002,Alain2006,Cheng2020}, GAGG(Ce) provides a 
non-hygroscopic alternative without intrinsic radioactivity, enabling simplified 
packaging and long-term operation. On the other hand, Silicon Photomultipliers (SiPMs) represent a state-of-the-art photosensor technology, offering significant advantages over conventional Photomultiplier Tubes (PMTs)~\cite{Brivitch2006,Jackson2015}. SiPMs are compact, operate at low bias voltages, and are insensitive to magnetic fields, making them ideal for constructing highly granular and robust detector arrays. 
The combination of GAGG(Ce) scintillators with SiPM readout enables the development of a compact $\gamma$-ray detector array with good energy and timing resolutions, high detection efficiency, and outstanding operational stability.
Recent studies conducted within the HYPATIA project at RIKEN have characterized the performance of large-volume GAGG(Ce) scintillators coupled to SiPMs for $\gamma$-ray spectroscopy~\cite{Cortes2026,HypatiaWeb}.

In this work, we present the development and performance evaluation of GAGG(Ce) detector modules equipped with SiPM readout. The detector module design is described in detail, and the performance of ten modules—including energy linearity, energy and timing resolution, and intrinsic full-energy peak detection efficiency—has been  evaluated using standard $\gamma$-ray sources and the $^{14}$N$(p,\gamma)^{15}$O reaction, covering a $\gamma$-ray energy range from about 0.3 to 8 MeV. The results demonstrate that the modules are a good solution for 
next-generation large-scale in-beam $\gamma$-ray spectrometers.

%%%%%%%%%%%%%%%%%%%%%%%%%%%%%%%%%%%%%%%%%%%%
\section{GAGG detector module design}

The design of the GAGG detector module is illustrated in Fig.\ref{fig:fig1} (a).
Each module incorporates four individual detectors, with each detector consisting of a 25 $\times$ 25 $\times$ 75 \si{\milli\metre^3} GAGG(Ce) crystal.
The GAGG(Ce) crystals are manufactured by China Electronics Technology Group Corporation, which has a high light yield of more than 55000 photons/MeV and an emission spectrum that peaks at 530 nm.
The crystal length of 75 mm is a trade-off between crystal cost and detection efficiency for high-energy $\gamma$ rays.
The crystal features a polished readout surface for optimal optical coupling to a Hamamatsu S14161-6050HS-04 SiPM ~\cite{hamamatsu_S14161_6050HS_04} via a 1-mm thick light guide.  
This SiPM has a low breakdown voltage of 38 V and exhibits a photon detection efficiency around 50\% at $\sim$450 nm, while maintaining an efficiency of $\sim$35\% at the emission peak of GAGG around 530 nm. It is assembled from a 4 $\times$ 4 array of sub-units, each with an effective photosensitive area of 6.0 $\times$ 6.0 mm$^{2}$, resulting in an overall dimension of 25 $\times$ 25 mm$^{2}$, thereby providing good geometrical matching with the GAGG crystal.
The remaining five surfaces of the crystal are frosted to enhance light collection efficiency via diffuse scattering. 
The crystal was wrapped with a 65 $\mu$m enhanced spectral reflector (ESR) film to serve as the reflective material. For light-tightness and additional reflection, the assembly was further encased in $\text{BaSO}_{4}$ with a thickness of about 230 $\mu$m.

\begin{figure*}
    \centering
    \includegraphics[width=\linewidth]{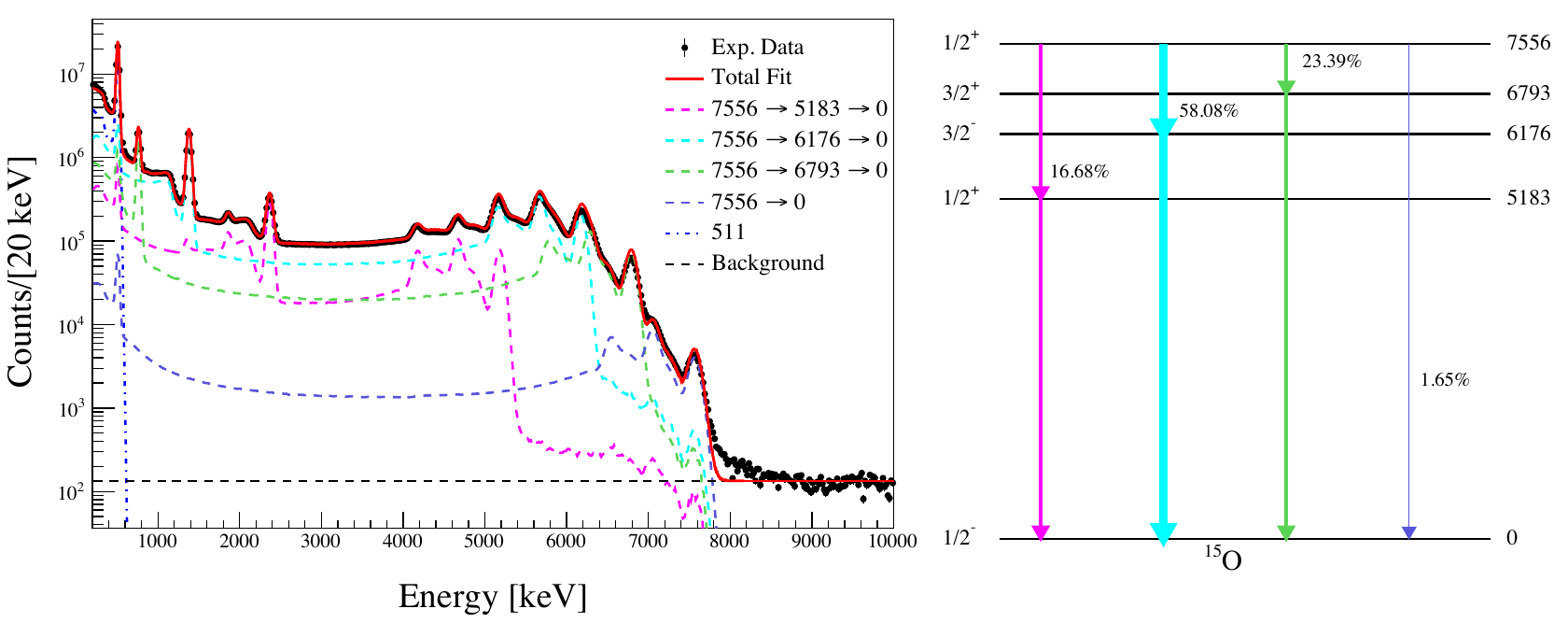}
    \caption{Energy spectrum obtained using the $^{14}$N($p,\gamma$)$^{15}$O reaction. The right panel shows the decay scheme of the $^{15}$O 7556 keV excited state, including its decay branching ratios. The spectrum is fitted with a sum of four response functions obtained from Geant4 simulations and a constant background to account for background from unresolved high-energy transitions. 
    Additional 511 keV $\gamma$ rays from positron annihilation originating from $\beta^+$ decay of $^{15}$O and unobserved $\gamma$ rays at higher energies are also considered in the fitting. The simulated spectra were convoluted with the measured energy resolution before comparison with experimental spectra.
    }
    \label{fig:gammafit}
\end{figure*}

For each detector, the signals from the 16 SiPMs are summed and subsequently amplified with a tunable gain ranging from $\sim$10 to $\sim$70, enabling adjustment of the energy dynamic range. The output signal polarity is positive. The front-end electronics can be powered by an external 12 V DC supply, which is converted to provide the required SiPM bias voltage (approximately 40 V) and the $\pm$3.3 V supply for the operational amplifier. The power consumption of a single detector module is approximately 400 mW.
To compensate for the temperature dependence of the SiPM breakdown voltage and the
resulting gain variation at a fixed bias voltage~\cite{Otte2017,Anfimov2021}, a temperature sensor is integrated onto the SiPM board.
The measured temperature is used by a feedback circuit to dynamically adjust the SiPM bias voltage so that the SiPM gain is stabilized.
The applied bias voltage is parameterized as a third-order polynomial of the temperature. The coefficients are determined by placing the module in a temperature-controlled chamber and measuring the spectrum of a radioactive source. These coefficients can be updated as required.
All the detector modules are connected to a control board communicated with a computer via an RS485 serial interface.

When dealing with hygroscopic scintillators such as NaI(Tl) or LaBr$_3$(Ce), hermetic encapsulation is required to prevent moisture degradation.
In contrast, GAGG(Ce) exhibits excellent non-hygroscopic properties, eliminating the need for complex moisture-proof packaging and allowing the use of lightweight carbon fiber encapsulation. 
The present detector module houses four GAGG(Ce) crystals within a thin carbon fiber enclosure with a wall thickness of only \SI{300}{\micro\meter}. This enclosure is glued to an aluminum block that sandwiches the crystals and readout boards securely inside. The position of the module can be mechanically fixed solely through connections to the aluminum block, which provides sufficient structural support for approximately a 1.5 kg assembly. This module design minimizes the material budget, enabling a compact configuration while maximizing the angular coverage of the detector array for in-beam $\gamma$-ray spectroscopy experiments.
%%%%%%%%%%%%%%%%%%%%%%%%%%%%%%%%%%%%%%%%%%%%

\section{Experimental setup and simulations}

The test experiment was conducted at the 350 kV accelerator facility at Hefei Institutes of Physics Science, Chinese Academy of Sciences
~\cite{hefei_CHEN2023168401} with $\sim$1 mA proton beam and TiN targets. 
The accumulated proton beam intensity was 3.95 C with a relative uncertainty of 1\%. 
The schematic layout of the experimental setup is illustrated in Fig.~\ref{fig:fig1} (b). A total of 10 detector modules were placed symmetrically around the target, covering 
polar angles of about $30^{\circ}$--$160^{\circ}$ and an azimuthal angular range of about $\pm 60^{\circ}$. 
The signals from each detector were fed into the DT2740 CAEN digitizer. Each channel of the digitizer digitizes the analog input with a 16-bit, 125 MS/s ADC.

High-energy $\gamma$-rays up to $\sim$8 MeV were produced using the $E_{p}$ = 278 keV resonance of the $^{14}$N($p$, $\gamma$)$^{15}$O reaction.
This reaction populates the 7556-keV excited state of $^{15}$O, which subsequently de-excites through several dominant $\gamma$-ray transitions, as shown in the right panel of Fig.~\ref{fig:gammafit}~\cite{Imbriani2005,Chen2025,NNDC-O15}.
The production yield of the 7556-keV excited state was $3.85\times10^{4}$/mC with a relative uncertainty of 2\%.
In addition, measurements with standard radioactive sources ($^{22}$Na, $^{60}$Co, 
and $^{152}$Eu) were performed to characterize the low-energy response, energy 
resolution, and timing performance of the detector modules. A natural $^{40}$K 
source prepared from a KCl sample was used to evaluate the long-term stability.

The full experimental setup of the in-beam test was modeled in a Geant4 simulation, including the GAGG(Ce) crystals, optical reflectors, carbon-fiber enclosures, lead shielding and the vacuum chamber. The measured detector positions and source geometry were implemented according to the experimental configuration. Monoenergetic $\gamma$ rays with energies corresponding to the observed transitions in the $^{14}$N$(p,\gamma)^{15}$O reaction or those from standard radioactive sources 
were generated isotropically from the target position or the corresponding source 
position. The simulated energy spectra were obtained by recording the energy deposition in 
the GAGG(Ce) crystals and were convoluted with the measured energy resolution to account for the detector response. The intrinsic full-energy peak (FEP) efficiency was determined as the ratio of the number of events within the full-energy peak to the total number of generated $\gamma$ rays. 
The experimental FEP yields were extracted by fitting the measured 
energy spectra with the simulated response functions, and the corresponding efficiencies were obtained by normalization to the number of incident $\gamma$ rays.
The simulation was performed using the FTFP\_BERT\_HP physics list combined with 
G4EmStandardPhysics\_option4 for electromagnetic interactions.

\section{Results}

\subsection{Energy Calibration, Linearity and Energy Resolution}

\begin{figure}
    \centering
    \includegraphics[width=\linewidth]{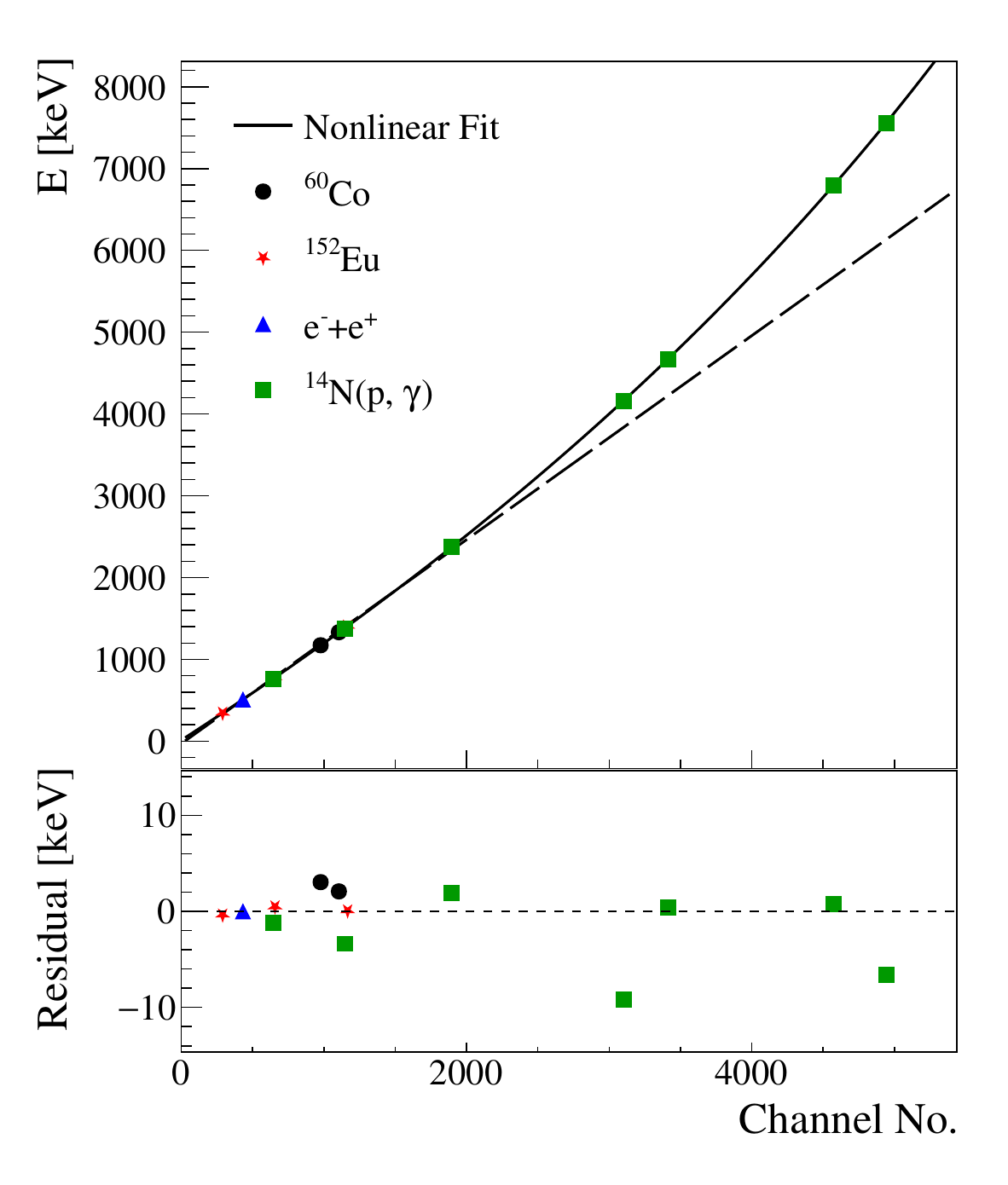}
    \caption{Energy calibration of the GAGG detectors. 
    The relationship between energy and channel is fitted using the exponential function in Eq.\ref{eq:sipm3} and the corresponding residual distribution is shown in the lower panel. The dashed line in the upper panel is included as a guide to the eye and represents a linear fit to the low-energy region below 2000 channels.}
    \label{fig:calib}
\end{figure}

\begin{figure}
    \centering
    \includegraphics[width=0.98\linewidth]
    {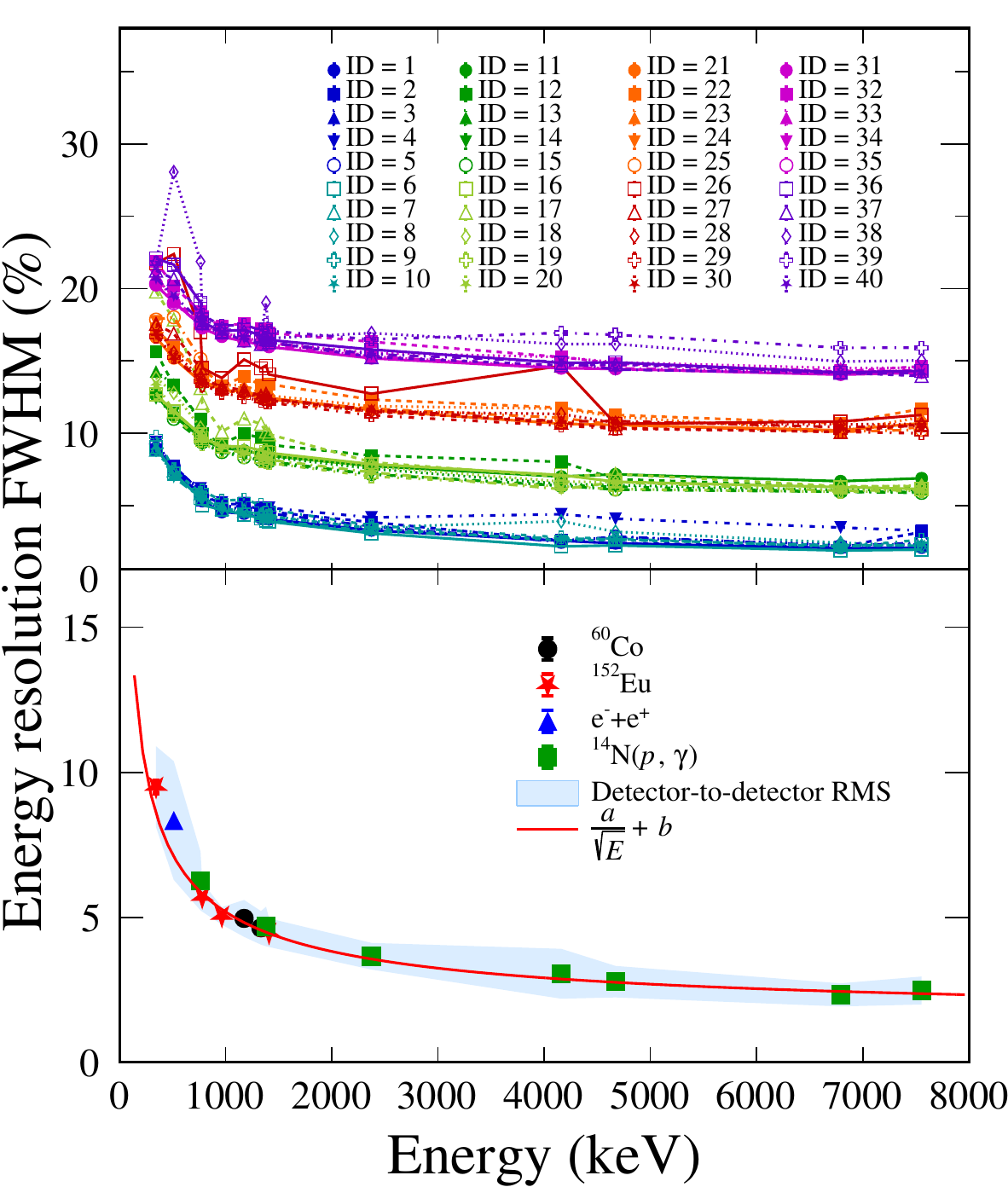}
    \caption{
    The energy resolutions of 40 GAGG detectors as a function of the $\gamma$-ray energies. For clarity, detectors are grouped according to their IDs (1–10, 11–20, 21–30, and 31–40). Vertical offsets of 0\%, 4\%, 8\%, and 12\% are applied to each group, respectively, for visualization purposes. The lower panel shows the average energy resolution, while the RMS variation among detectors is indicated by the light blue band. The energy dependence of the energy resolution can be well approximated by $\frac{a}{\sqrt{E}} + b$.
    }
    \label{fig:resolution}
\end{figure}

The obtained energy spectrum from the $^{14}$N($p,\gamma$)$^{15}$O reaction is shown 
in Fig.~\ref{fig:gammafit}. The experimental spectrum was fitted using the simulated 
response functions described above, with the measured energy resolutions taken into 
account. The energy calibration and energy resolutions were extracted as described below.

By comparing the reconstructed energy and channel values, a clear non-linear response was observed, especially at high $\gamma$-ray energies as shown in upper panel of Fig.~\ref{fig:calib}. This behavior is attributed to the saturation effect of the SiPM 
microcells. In a SiPM, each microcell can detect a single photon at a time. Due to the finite number of available microcells, the SiPM output exhibits a saturation effect as the incident photon flux increases, resulting in a non-linear relationship between the detected signal amplitude and the deposited energy. In cases where the scintillation decay time of the crystal is comparable to the SiPM recovery time ($\sim 100\text{ ns}$), the following empirical formula can be employed to correct the SiPM non-linearity~\cite{hamamatsu_lineary}:
\begin{equation}\label{eq:sipm1}
N_{\text{fire}} = N_{\text{eff}}( 1 - e^{-N_{\text{pd}}/N_{\text{eff}}}),
\end{equation}
\begin{equation}\label{eq:sipm2}
N_{\text{fire}} = \frac{CH}{CH_{\text{s}}}.
\end{equation}
In these expressions, $N_{\text{fire}}$ denotes the number of fired microcells per event, which is defined as the ratio of the charge integration peak position ($CH$) to the single-photon peak position ($CH_{\text{s}}$). 
$N_{\text{pd}}$ represents the expected number of detected photons, determined by the product of the photon detection efficiency ($\epsilon$) and the number of scintillation photons reaching the SiPM ($N_{\text{in}}$) that is proportional to the deposited energy $E$: $N_{\text{in}} = f \cdot E$. 
Therefore, $N_{\text{pd}}$ = $\epsilon$ $\cdot f \cdot E = k \cdot E$.
$N_{\text{eff}}$ is the effective number of available microcells. When the SiPM recovery time is 
shorter than the scintillation decay time, partial pixel recharge allows multiple firings within a single event, and therefore $N_{\text{eff}}$ typically exceeds the physical number of microcells ($N_{\text{pix}}$). 
By combining Eqs. (1) and (2), we obtain:
\begin{equation}\label{eq:sipm3}
CH = N_{\text{eff}} \left( 1 - e^{-k \cdot E / N_{\text{eff}}} \right) \cdot CH_{\text{s}}.
\end{equation}

We therefore adopted a three-parameter function to describe the channel-energy relation
$Channel = a \times (1 - exp(-b \times E)) + c$. 
To ensure calibration validity across a broad energy range, the parameters $a, b$ and $c$ were determined   
using the full energy peaks of the following $\gamma$-rays sources: 
1173 keV and 1332 keV from $^{60}$Co; 
334 keV, 779 keV and 1408 keV from $^{152}$Eu;
511 keV from positron annihilation;
763 keV, 1380 keV, 2373 keV and 6791 keV from the ($p$, $\gamma$) reaction; 
4671 keV and 4160 keV, single- and double-escape peaks associated with the 5182 keV transition in the ($p$, $\gamma$) reaction. 
The peak centroids were extracted via either a pure Gaussian fit or a Gaussian fit incorporating an exponential background in the peak region, predominantly arising from Compton scattering of higher-energy transitions. The experimental data and corresponding fitting result for a representative detector are illustrated in Fig.~\ref{fig:calib}. The residual distributions demonstrate that the energy deviations are confined within 10 keV. 

In contrast to conventional scintillators such as PMT-readout NaI(Tl), which typically exhibit near-linear light output, the present SiPM-based readout GAGG detectors clearly show that the microcell saturation becomes significant at high scintillation light yield. As a result, calibration parameters determined from low-energy radioactive sources can lead to a deviation of more than 1 MeV when extrapolated to the high-energy region. This makes it particularly important to perform energy calibration in a wide dynamic range for large GAGG(Ce) crystals intended for in-beam $\gamma$-ray spectroscopy at high incident beam energies, where strong Lorentz boosts can extend the detected $\gamma$-ray energies of 10 MeV or beyond.

The energy resolution, defined as the full width at half maximum (FWHM) divided by the peak centroid, 
was extracted using the same fitting procedure as for the peak centroid determination in the peak region (Gaussian or Gaussian plus exponential background).
Systematic uncertainties associated with the background parameterization were estimated to be on the order of 10\%.
The extracted energy resolution of individual detectors as a function of the $\gamma$-ray energies are shown in Fig.~\ref{fig:resolution}.
The typical energy resolution around 1 MeV is ~5\% and increases to $\sim$2\% at 7 MeV.
The energy dependence of the resolution can be well parameterized by
$a/\sqrt{E}+b$, except for detector IDs 26 and 38, which show larger deviations
possibly due to variations in crystal quality or light collection. The first term
represents the contribution from photon statistics, while the constant term is mainly associated with residual SiPM nonlinearity.
The present GAGG(Ce)-SiPM modules achieve improved energy resolution compared with commonly used large-volume PMT-readout NaI(Tl) detectors ~\cite{Takeuchi2014} or APD-readout CsI detectors~\cite{Cortina2014}, which typically achieve resolutions worse than 5\% at 1 MeV.
However, the obtained energy resolution is
about a factor of two worse than that achieved with large-volume
LaBr$_3$(Ce) scintillators~\cite{Weisshaar2008,Aogaki2023,Wei2025}, owing to the higher light yield of LaBr$_3$(Ce).

 \begin{figure*}[t!]
   \setlength{\abovecaptionskip}{0pt}
   \setlength{\belowcaptionskip}{0pt}
   \centering
   \includegraphics[width=\textwidth,height=17cm]
     {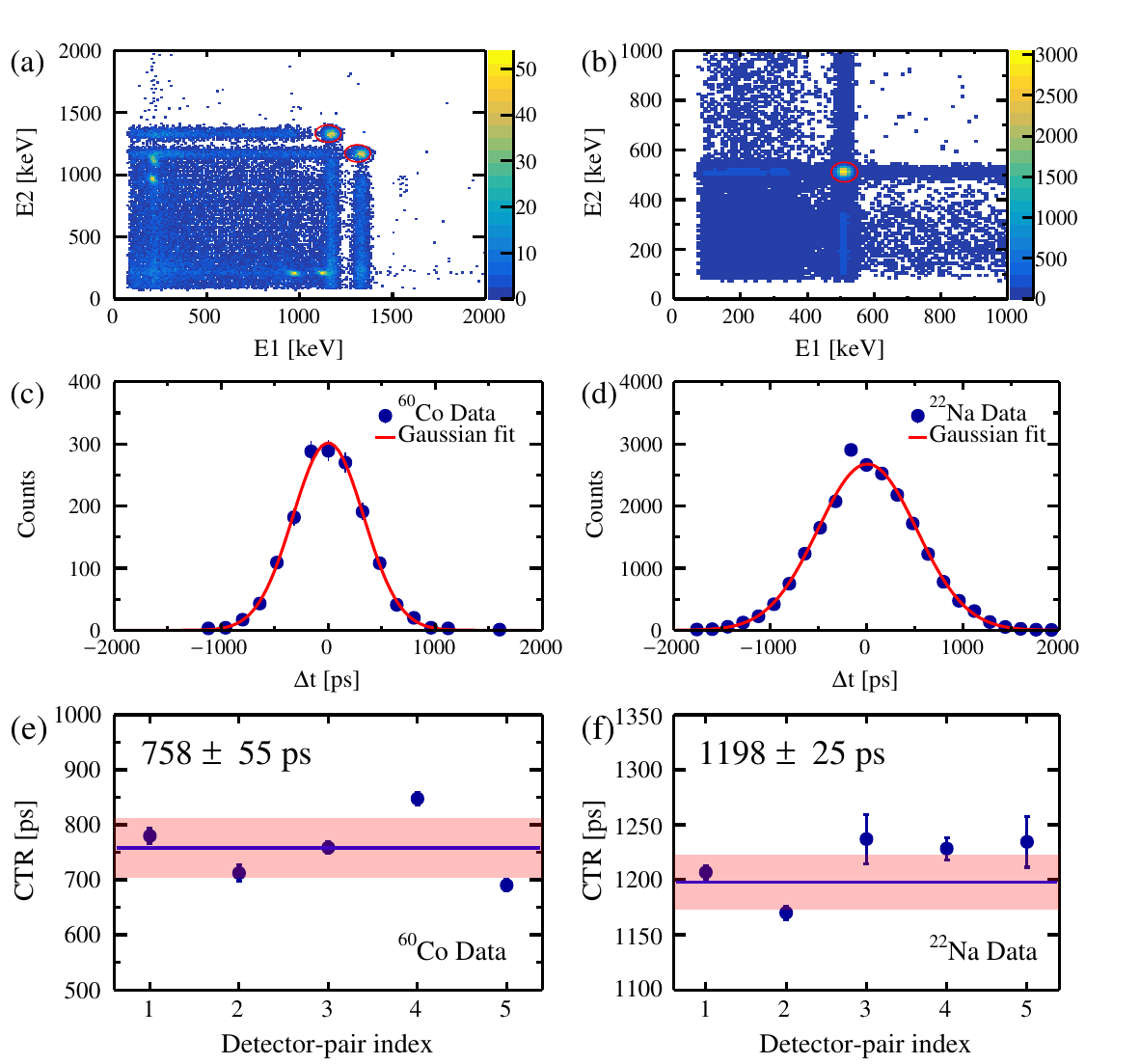}
    \caption{Timing resolution characterization employing 
$^{60}$Co and $^{22}$Na radioactive sources. (a), (b) Energy correlations for double-coincidence events, where the events in the highlighted circles correspond to the 1173 and 1332 keV cascade gamma rays from $^{60}$Co and the 511 keV annihilation photons from 
$^{22}$Na. (c), (d) Time-difference distributions gated on the full-energy peaks. Red lines show fitted results with Gaussian functions. (e), (f) Derived coincidence time resolutions for five representative pairs of detectors. The blue lines and red bands represent the mean values and RMS spreads, respectively.}
    \label{fig:time_res}
\end{figure*}

\subsection{Timing resolution}

Timing resolution plays an important role in in-beam $\gamma$-ray spectroscopy, where it is essential for suppressing background events arising from 
random coincidences and delayed atomic or nuclear processes,
which can extend up to the 10 ns timescale~\cite{Weisshaar2008}. 
In add-back mode, where signals from neighboring $\gamma$-ray detectors within a cluster are combined, events are correlated within a coincidence time window typically of the order of 1–2 ns. 
A good timing resolution better than 1 ns is required to enable further narrowing of the coincidence window and suppression of background, thereby improving the add-back performance and peak-to-background ratio.

Here, the timing resolution of the GAGG detectors was measured using either a $^{60}$Co or a $^{22}$Na source placed at the center of the detector array at a distance of approximately 5 cm from the 25$\times$75 mm$^{2}$ entrance surface of the detectors.
Since $^{60}$Co emits two $\gamma$ rays in cascade and $^{22}$Na produces back-to-back 511 keV $\gamma$ rays, the time difference between signals detected in two detectors was used to extract the coincidence timing resolutions (CTR) and the single-detector timing resolutions (STR).
Instead of the DT2740 CAEN digitizer, the DT5730 CAEN digitizer, featuring a higher sampling rate of 500 \si{ MS/s}, was used for data acquisition. Leading-edge discriminator (LED) mode was adopted for the time measurement. 
The DT5730 digitizer has only eight input channels; therefore, due to the limitation of available channels,
two independent measurements were performed that cover 12 detectors.
Coincidence events were clearly observed as shown in panel (a) and (b) of Fig.~\ref{fig:time_res}.
For the $^{22}$Na measurement, detector pairs placed symmetrically on opposite sides of the $\gamma$-ray source have been selected.
The time-difference spectra were then obtained by applying $\pm1\sigma$ energy window around the photopeak in both detectors, in order to suppress the Compton scattering background and to reduce amplitude-dependent time walk associated with the LED timing method.

The CTR were determined by fitting the time-difference spectra with a Gaussian function, where the CTR is defined as the FWHM of the peak. 
The typical spectra are shown in panels (c) and (d) in Fig.~\ref{fig:time_res}.
For the $^{60}$Co and $^{22}$Na measurements, the CTR values obtained from five representative detector pairs are summarized in panels (e) and (f), respectively.
The mean CTR values are 758(5) ps for $^{60}$Co and 1198(4) ps for $^{22}$Na, with RMS spreads of 55 ps and 25 ps.

For the $^{60}$Co measurement, the large number of valid coincidence combinations enables the extraction of the STR without assuming identical detector performance.
For each detector pair, the measured CTR satisfies $R_{ij}^{2}=R_i^{2}+R_j^{2}$,
where $R_i$ and $R_j$ denote the STR for detector $i$ and detector $j$ respectively.
The resulting system of equations was solved using a least-squares method considering all measured detector pairs.
The extracted STR range from 459 ps to 617 ps, with an average value of 524 ps and an RMS spread of 50 ps.

For the $^{22}$Na measurement,
only detector pairs with opposite geometrical positions can provide coincidence events, preventing the extraction of STR using the global fitting method.
Assuming similar timing performance among the detectors, the average STR is estimated as
$1198/\sqrt{2}=847$ ps.
For comparison, the estimate for the $^{60}$Co measurement gives 
$758/\sqrt{2}=536$ ps, 
which is reasonably consistent with the value obtained from the least-squares extraction.

The observed ratio of the two resolutions, $847/536 \approx 1.58$, is reasonably consistent with the scaling of 
$\sqrt{E_{\gamma}(^{60}\mathrm{Co})/E_{\gamma}(^{22}\mathrm{Na})}\approx1.52-1.61$.
This indicates that the timing resolution approximately follows a $1/\sqrt{E}$ dependence. 
It should be noted that, in this estimation, the contribution from the time-of-flight spread due to the finite source and detector dimensions has not been subtracted.
Considering the source–detector distance ($\sim$ 5 cm) and detector size, this contribution is on the order of $\sim$40 ps (FWHM).
The DT5730 waveform digitizer provides digital timing extraction with a fine interpolation precision of 2 ps.
Therefore, the contribution from the waveform digitizer is also expected to be small compared with the detector response.
Since the narrow energy gate suppresses the amplitude-dependent time walk, the remaining timing spread is
mainly determined by the detector response, including uncertainty of the $\gamma$-ray interaction position inside the crystal,
timing jitter arising from scintillation photon statistics, light transport time spread and SiPM response fluctuations.

Sub-100 ps timing resolution has been demonstrated using small-volume GAGG(Ce) crystals with SiPM readout~\cite{Martinazzoli2021}. 
The present detector employs a much larger GAGG(Ce) crystal (25 $\times$ 25 $\times$ 75 mm$^{3}$), for which optical photon transport becomes important. The longer optical path increases the probability of multiple internal reflections and absorption, reducing the light transfer efficiency and broadening the distribution of photon arrival times at the SiPM. Consequently, both the energy resolution and timing resolution are expected to deteriorate~\cite{Gundacker2014}.
To our knowledge, the present result represents one of the best timing performances reported for large-volume GAGG(Ce) scintillators. 
The sub-nanosecond timing resolutions significantly outperform conventional NaI- and CsI-based detectors, which typically exhibit nanosecond-scale timing resolution. 
Nevertheless, the obtained timing resolution is about a factor of two worse than that achieved with large-volume LaBr$_3$(Ce)~\cite{Weisshaar2008, Aogaki2023, Wei2025} and CeBr$_3$~\cite{Qin2020}, which require hermetic encapsulation due to their hygroscopic nature.

\begin{figure}
    \centering
    \includegraphics[width=1\linewidth]{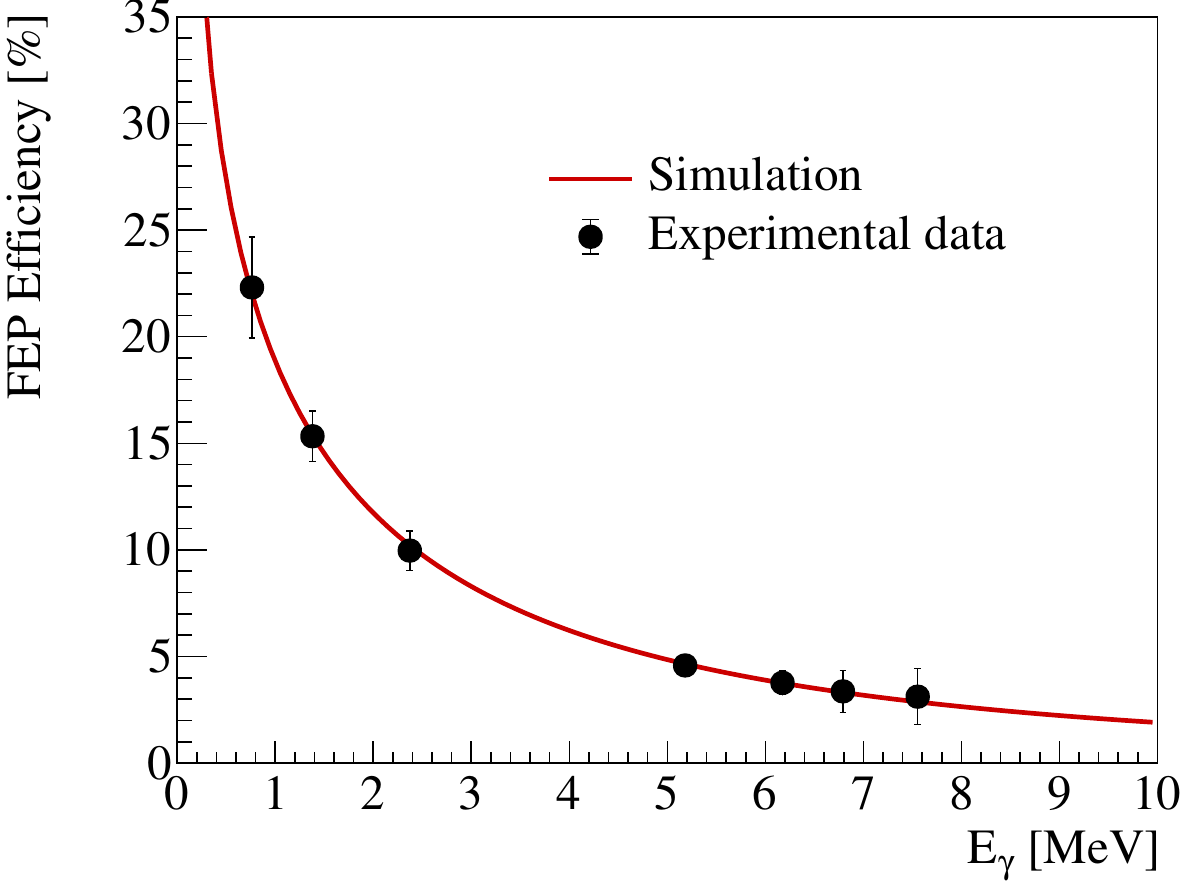}
    \caption{Intrinsic full-energy peak detection efficiency of the detector array consisting of 10 GAGG detector modules. The experimental data include 7 $\gamma$-ray transitions of the $^{15}$O 7556-keV excited state.
    The error bars mainly reflect the uncertainty in the extracted peak intensities, estimated by varying the fitting range of the $\gamma$-ray spectra shown in Fig.~3. The red solid line represents the efficiency obtained from Geant4 simulations.}
    \label{fig:eff}
\end{figure}

\subsection{Intrinsic FEP Efficiencies}

The intrinsic FEP efficiency of the entire array was further evaluated across an energy range from 0.76 MeV to 7.55 MeV, utilizing the $^{14}$N($p$,$\gamma$)$^{15}$O reaction. 
Fig. ~\ref{fig:eff} compares the measured FEP efficiencies with the corresponding Geant4 simulations using the same detector geometry and source configuration. 
Good agreement is observed over the entire investigated energy range.
The measured efficiencies decrease from $\sim$22\% at 0.76 MeV to $\sim$3\% at 7.55 MeV, following the expected energy dependence of the full-energy peak efficiency 
due to the reduced photoelectric interaction probability and the increasing contribution of Compton scattering at higher $\gamma$-ray energies. 
The simulated efficiency curve was parameterized using a logarithmic polynomial function:

\begin{equation}
\epsilon(E_\gamma)=
\exp\left(a_0+a_1\ln(E_\gamma)+a_2\ln^2(E_\gamma)+a_3\ln^3(E_\gamma)\right),
\end{equation}

\noindent
which provides a convenient description of the detector response and can be used for efficiency correction in future in-beam $\gamma$-ray spectroscopy experiments.

\begin{figure}
    \centering
    \includegraphics[width=1\linewidth]{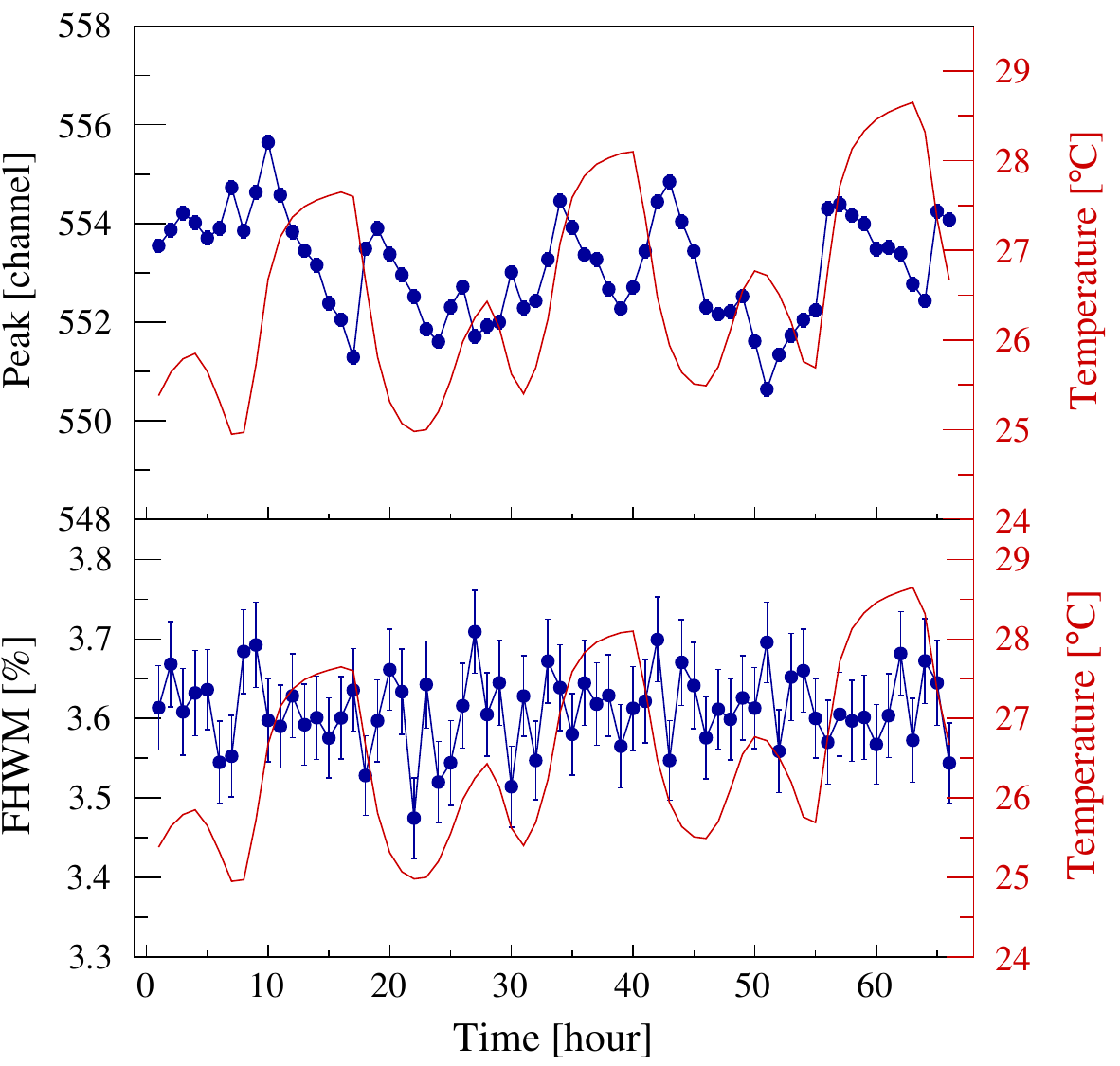}
    \caption{Evolution of peak position and energy resolution as a function of measurement time using $^{40}$K for one typical detector. The red solid lines show the change of the temperature measured by the sensor integrated on the SiPM board.}
    \label{fig:stability}
\end{figure}

\subsection{Long-Term Stability}

Finally, we tested the long-term stability of the detector using a natural $^{40}$K source ($E_{\gamma} = 1461$ keV) at ambient temperature in air without active cooling. The detector was continuously operated for about 7 days. 
The adopted S14161-6050HS-04 SiPM has a temperature coefficient of 34 mV/$^\circ$C~\cite{hamamatsu_S14161_6050HS_04},
which results in a gain variation of approximately 1.7\%/$^\circ$C when operated at a fixed bias voltage of 40 V.
With temperature compensation, this gain variation is strongly suppressed.
Figure ~\ref{fig:stability} shows the evolution of the peak position and energy resolution over a representative 3-day period, 
together with the corresponding temperature variation measured by the sensor integrated on the SiPM board.
The maximum peak shift is about 5 channels, corresponding to an energy shift of approximately 13 keV, or a maximum relative peak-position drift of $\sim 0.9\%$, during a temperature variation of about 4~$^\circ$C. 
Despite the temperature fluctuation of several degrees, the energy resolution remains stable within the measurement period, with no observable correlation between the FWHM and temperature.
These results demonstrate that the
temperature compensation feedback circuit effectively suppresses the SiPM gain variation
and enables stable operation of the GAGG(Ce) modules for long-duration $\gamma$-ray spectroscopy experiments.

\section{Summary}

In summary, we have developed a compact large-volume GAGG(Ce) detector module with SiPM readout optimized for $\gamma$-ray spectroscopy.
To minimize $\gamma$-ray attenuation in the passive material while maintaining the mechanical robustness,
each module was designed to house four large-volume GAGG(Ce) crystals within a thin carbon-fiber enclosure.
Ten modules were evaluated using standard $\gamma$-ray sources and the $^{14}$N$(p,\gamma)^{15}$O reaction over an energy range of about 0.3–8 MeV. The energy response exhibits non-linearity at higher energies and can be well parameterized by accounting for SiPM saturation effects. The detectors show good energy resolution, FWHM below $\sim$5\% at 1 MeV, sub-nanosecond timing resolution, and intrinsic full-energy peak efficiencies in good agreement with Geant4 simulations. 
The present large-volume GAGG(Ce)-SiPM modules provide a promising solution for future large-scale $\gamma$-ray detector arrays planned for next-generation high-efficiency, large-solid-angle in-beam $\gamma$-ray experiments at RIKEN~\cite{HypatiaWeb,Cortes2026} and HIAF~\cite{Li_2024}.

\section*{Acknowledgments}
We are very grateful to the staff of Hefei Institutes of Physical Science for providing the stable and high-intensity proton beam.
This work has been supported by the Fundamental Research Funds for the Central Universities and the National Natural Science Foundation of China (Grant No. 12575121, 12325506, 12550008).
H.N.L., Y.L.J., T.Q.L., Y.L.L., and Y.Y.Y. acknowledge the support from the National Key R\&D Program of China (Grant No. 2023YFA1606403, 2024YFE0102800), the National Natural Science Foundation of China (Grant No. 12375111, 12361141822), and the Fundamental Research Funds for the Central Universities of China.
We thank L. Wang, X. Xiao, W. Hu, T.T. Qu, P. Doornenbal and M.L. Cortes for the valuable discussions on the design of the detector module.

%\clearpage

\bibliographystyle{unsrt} 
\bibliography{references}  

\end{document}